# Fall Detection using Knowledge Distillation Based LSTM for Offline Embedded and Low Power Devices

Hannah Zhou[#1], Allison Chen[2], Celine Buer[#3], Kayleen Tang[4]

Lauryn Gong[5], Emily Chen[6], Zhiqi Liu[7], Jianbin Tang[8]

[1#]*Holliston High School, Holliston, MA, USA*
[1]hyz4@cornell.edu, [3]celinebuer19@gmail.com

[2*]*Hopkinton High School, MA, USA*
[2]allytotheson@gmail.com

[4*]*Deerfield Academy, MA, USA*
[4]Kayleentang718@gmail.com

[5*]*Dover Sherborn High School, MA, USA*
[5]lauryn.gong@gmail.com

[6*]*Newton North High School, MA, USA*
[6]emilyaychen@gmail.com

[7*]*Boston Dynamics, MA, USA*
[7]zliu@bostongdynamics.com

[8*]*IBM Research, Australia*
[8]jbtang@au1.ibm.com

*Abstract—*
**This paper presents a cost-effective, low-power approach to unintentional fall detection using knowledge distillation based LSTM (Long Short-Term Memory) models to significantly improve accuracy. With a primary focus on analyzing time-series data collected from various sensors, the solution offers real-time detection capabilities, ensuring prompt and reliable identification of falls. The authors investigate fall detection models that are based on different sensors, comparing their accuracy rates and performance. Furthermore, they employ the technique of knowledge distillation to enhance the models' precision, resulting in refined accurate configurations that consume lower power. As a result, this proposed solution presents a compelling avenue for the development of energy-efficient fall detection systems for future advancements in this critical domain.**

## I. Introduction

Falls are a significant public health concern, particularly for seniors aged 65 years and older, and are the leading cause of injury-related deaths in this population [1]. These injuries not only impact the lives of older adults but also their loved ones. Real time fall detections devices can help prevent injuries by allowing a way to get quick assistance, while also improving quality of life and peace of mind for caregivers. Such devices need to be low-cost, low power, simple, and able to identify motion patterns with accuracy.

This paper introduces a fall detection solution for cost-effective and low-power embedded devices. The system integrates LSTM models based on knowledge distillation to enhance the detection accuracy. The proposed approach utilizes time-series data collected from multiple sensors to recognize fall patterns in real-time. To train the model, online datasets are used and the device is designed for everyday usage. Experimental results demonstrate that the proposed solution achieves high accuracy in detecting falls, while maintaining low power configurations.

The contribution of this work is as follows. Firstly, it analyzes the impact of different sensor types on fall detection model accuracy. Existing studies [2,3,4,5,6,7,8,9] have used various sensors, such as accelerometers, gyroscopes, and barometers, but our experiments clarify the influence of each sensor or combination of sensors. Secondly, knowledge distillation is used to enhance the accuracy of single-sensor models by transferring knowledge from more complex models. Finally, we identify the most low-power and high-accuracy configurations by considering both model accuracy and power efficiency. This approach facilitates the development of fall detection systems that are both effective and energy-efficient.

The paper is organized as follows. Section II surveys prior research on deep learning algorithm based fall detection. Section III outlines the dataset and LSTM model used in this study. Models that use different sensors are evaluated. Section IV discusses the techniques that are used to balance model accuracy and complexity. Section V proposes design space and method to develop low power and high accurate fall detection configuration, followed by concluding remarks in Section VI.

## II. Related work

Sensor based fall detection that use deep learning models can be categorized as collaborative device/server systems and offline devices.

The collaborative device/server systems [2,3] collect sensor data on the device, send these data to associated servers, and get the inference results from the server to the device. Such configuration supports more complicated models on the server and thus significantly improves the prediction accuracy. This architecture, however, requires additional network and server along with the fall detection devices. It is hard to use, not cost effective, or power efficient.

The offline devices system [4,5,6,7,8,9] deploy models on devices to predict human activity types and detect fall. The models are trained using labeled data, which is collected from simulated daily activities and manually labeled with corresponding activity names by testers. Research on offline device systems focus either on model accuracy or energy efficiency.

On-device training focuses on machine learning models that can be trained directly on devices without the need for transmitting data to a centralized server. Google introduced Federated Learning[10] for models trained from user interaction with mobile devices which allows for mobile phones to collaboratively learn a shared prediction model, keeping all the training data on the device, and separating the need to store data in the cloud from the ability to do machine learning. Researchers also proposed TinyBERT[11], a technique that compresses large-scale pre-trained language models into smaller models suitable for on-device training. By distilling BERT's knowledge, TinyBERT aims to enhance natural language understanding. The transfer of knowledge is achieved through a two-stage learning framework specific to TinyBERT. This framework encompasses Transformer distillation, enabling the smaller model to capture both general-domain and task-specific knowledge from BERT.

To the authors' best knowledge, there is no study to choose sensors and improve model accuracy to support offline low power devices.

III. DATASET AND MODELING

Our study introduces a novel approach in fall detection by leveraging a public dataset to construct deep-learning models with diverse sensor data configurations.

*DataSet*
We are using an open dataset called FallAllD [4]. It is a record of human daily activities and falls simulated by 15 subjects wearing sensors on neck, waist, and wrist. After careful consideration, we decided to use data from the wrist device because of its suitability for continuous and non-intrusive monitoring.

*Modeling*
To achieve precise and reliable fall detection, we incorporated state-of-the-art techniques explored in previous works [6,7,8]. However, we adjusted and improved these techniques to achieve a higher accuracy. The choice of sensor type and sensor placement are crucial in determining the method of fall detection. Raw data is collected by the sensors to capture regular movements. Algorithms are then used to extract features from the sensor data, such as the acceleration magnitude and direction, body angle, and movement speed. With machine learning algorithms, these features are classified into fall or non-fall events. Repeatedly training the model using a falling motion dataset can improve model accuracy.

To detect a fall, recognizing the sequence of movements is crucial. To differentiate between normal activities and falls in a short amount of time, we use a temporal sequence of activities [12]. A long short term memory (LSTM) model is efficient with long-term dependencies through the use of an additional memory gate. The LSTM layers developed in this study are depicted in Figure 1.

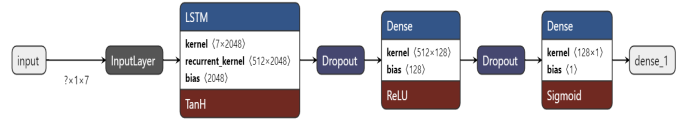

Fig 1. LSTM Network Layers. The model has 512 neurons in the LSTM layer, 128 in the hidden layer, and 1 in the output layer.

The initial Dense layer uses a Rectified Linear Unit (ReLU) Activation Function for analyzing intricate patterns while maintaining computational efficiency. Another Dense layer using a Sigmoid Activation Function maps the output of the model into a binary value of 0 or 1, representing label "Not Fall" and "Fall" labels, respectively. This binary classification is then used in the Binary Cross Entropy Loss Function.

| Sensor(s) Used | Model Accuracy |
|---|---|
| A, B, G | 93.55% |
| A, G | 82.37% |
| B, G | 89.43% |
| A B | 92.28% |
| A | 80.39% |
| G | 79.18% |
| B | 88.09% |

Table 1. Accuracy of models using different sensor configurations. A: accelerometer, B: barometer, G: Gyroscope.

*Result discussion*
As observed in Table 1, the model scores its highest accuracy when using all three feature vectors: accelerometer, barometer, and gyroscope. While A, G, and B, when used alone, have accuracies below 90%, Model ABG earns 93.55%.

In FallAllD, the LSTM model accuracy is 87.18%. The higher accuracy in our models is likely due to two noticeable differences. First, we categorize different activities into fall and no-fall, thus reducing the output dimension. Second, though we have experimented with several feature engineering techniques to preprocess the sensor data such as Fast Fourier Transform and moving average, none of these techniques achieve better prediction accuracy. So in the final model, we just use the raw sensor readings.

These results reveal that not all sensors contribute to fall detection the same. Barometers appear to be more effective at predicting falls, while the gyroscope is the least. However, since this dataset is from a fixed sensor configuration, in terms of sensor specifications and sampling frequency, we anticipate that devices with different configurations may produce other outcomes in practice.

## IV. NEW MODELING METHODS TOWARDS LOW POWER DESIGN

As observed in Table 1, accelerometers and barometers together predict nearly as accurately as all three sensors used together. Such prediction parity justify design choices of not using gyroscopes to reduce power consumption and cost, without significant impact on the prediction accuracy.

Since accuracy and configuration differences open opportunities for many design choices that aim for a variety of product goals, such as power efficiency, cost, accuracy, and reliability. We thus provide new methods to support these configurations.

This section investigates a knowledge distillation based process to design a simple yet highly accurate model.

### A. Knowledge Distillation

Knowledge distillation (KD) [13] is a technique in deep learning where the behavior of a larger and more complex model is transferred to a smaller model, thus computationally simpler model through learning and validation. It's been successfully used in compressing large models [14,15,16] as well as on resource constrained devices [17] and demonstrated to be effective in improving accuracy of smaller models used in low-power devices.

As the KD model is smaller, it is easier to deploy onto a Raspberry Pi 4 than the parent LSTM model is. Furthermore, considering the Raspberry Pi 4 is powered by a thin portable charger, maximizing the battery life is important, as it is inconvenient to keep recharging the battery over short intervals in the day.

We are developing and comparing the prediction accuracy and model complexity of three models: The parent LSTM model, a small model with the same number of layers but half the neurons, and a KD model with the same number of layers, half the neurons of the parent model, and a configurable number of sensors. For example, the KD model A uses only accelerometer data, removing gyroscope and barometer data from the model.

We aim to see if the KD Models can have similar prediction accuracy with fewer sensors, thus reduced cost and power consumption.

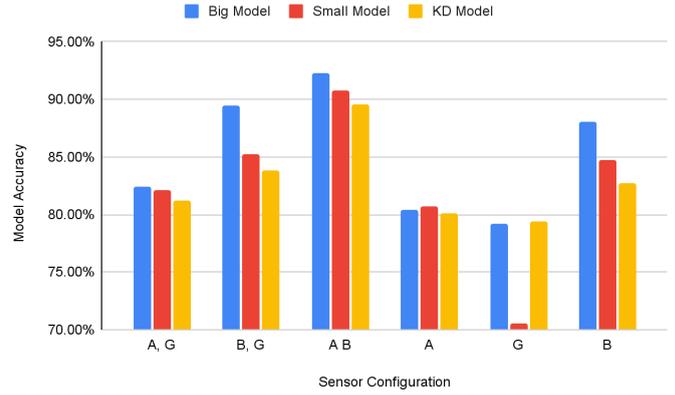

Fig 2. Model Accuracy comparison. The x axis is the sensor type. The y axis is the model prediction accuracy. Accuracy from the three models using different sensors are plotted.

As observed in Figure 2, both the small and KD models have lower accuracy than the big model. This is expected. The KD models perform very closely as the small models. In fact, the prediction accuracy differences between the small model and KD model is less than 2%. The prediction accuracy differences between the KD model and the big model are less than 6%. Considering the KD models are smaller than the big model and use fewer features, these results are very promising for future low power device development.

If we want to use the KD Model with the highest accuracy, KD Model AB provides the best outcomes with an accuracy of 89.52%. It is only 2.76% lower than the parent LSTM Model. In addition, we only use 2 of the sensors, so it consumes lower power than the parent model.

We have deployed KD models on a Raspberry Pi 4 and benchmarked the inference latency. The inference latency of 20 samples of accelerometer readings is less than 20ms. This demonstrates that KD models can significantly reduce the computation complexity and accelerate model development for low power devices.

## V. FUTURE WORK

To support low power configurations, we extend the above fall detection process by employing knowledge distillation. This involves training simpler models that use fewer sensors and selecting sensors and models with the highest accuracy and lowest power consumption. This process is illustrated in Figure 3.

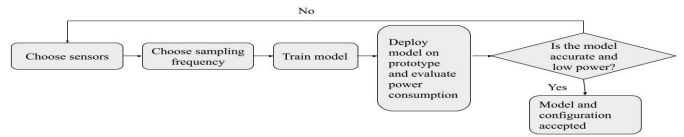

Fig 3. Sensor and Model Selection Process. In each iteration, sensors are chosen based on their contributions to prediction accuracy. Models that use these sensors are trained and evaluated on the prototype system. If the power consumption from model prediction and sensor sampling is the lowest, this model is used for the final product.

Low power design takes into account the energy consumption of controllers and sensors. Controller power depends on model complexity, while sensor power relies on its sampling rate as well as operating voltage and current.

Table 2 lists three sensor specifications. Notably, the accelerometer sensor has the lowest operating current, making it the most energy-efficient when other factors are held constant. Consequently, a configuration solely based on an accelerometer offers the potential for the most cost-effective and low power device. Nonetheless, for precise and dependable fall detection, both the sampling frequency and model accuracy are crucial. With these constraints in mind, the design space depicted in Figure 3 can be utilized to consistently evaluate the optimal configuration. We are conducting a study on this topic.

| Sensor Type | Model | Normal Operating Current |
|---|---|---|
| Accelerometer | MPU 6500 [18] | 450uA |
| Gyroscope | MPU 6500 [18] | 0.6mA |
| Barometer | BMP280 [19] | 3.2mA |

Table 2. Operating current of different sensor types.

## VI. Discussions and Conclusions

Designing a real time and accurate fall detection low power device faces many challenges. This study provides methods and experimental results to understand how different sensors contribute to fall detection models, the range of prediction accuracy among different model configurations, and directions to balance prediction accuracy and power efficiency.

## VII. Software and Implementation Details

The LSTM and KD models are implemented as Jupyter Notebook. They are available for review at https://github.com/Assistive-Technology-Create-Team/plumshum.github.io.

## VIII. Acknowledgement

We are grateful to the MIT Beaver Works Summer Institute for organizing the CRE[AT]E Assistive Technology Challenge, providing an exceptional single platform for the high school student authors of this paper, to apply coding and maker skills in a real-world assistive-technology project that helps community members with disabilities.

We express our sincere gratitude to the anonymous reviewers for their invaluable feedback and constructive suggestions, which greatly contributed to the improvement of this paper.